\documentclass[12pt]{article}
\usepackage{graphics}
\usepackage{amssymb}
\usepackage{epsfig,graphicx}
\usepackage{color}
\usepackage{amsmath}
\usepackage{amsthm}
\usepackage{amsopn}
\usepackage{subfigure}
\def\uno{\mbox{1 \kern-.59em {\rm l}}}

\def\be{\begin{equation}}
	\def\ee{\end{equation}}
\def\bea{\begin{eqnarray}}
	\def\eea{\end{eqnarray}}

\begin{document}
\begin{center}
{\bf{\large  Gravitational Effects on Quantum Coherence in Neutrino Oscillation
}}
\vskip 4em
{\bf M. M. Ettefaghi}\footnote{mettefaghi@qom.ac.ir}
 \vskip 1em
 Department of Physics, University of Qom, Qom 371614-6611,
Iran
\vskip 2em
{\bf R. Ramezani Arani}\footnote{ ramarani@kashanu.ac.ir }, and {\bf Z. S. Tabatabaei Lotfi}\footnote{ astrozat@gmail.com}	
\vskip 1em
University of Kashan, Km 6 Ravand Road, Kashan 87317-51167, Iran

 \end{center}
 
  \vspace*{1.9cm}
\begin{abstract}
In this paper, we investigate the quantum coherence for two flavor neutrinos propagating in a Schwarzschild metric. In fact, this issue is explored both qualitatively via calculating the parameter $K_{3}$ in Leggett-Garg inequality (LGI) and also quantitatively by evaluating the $l_{1}$-norm, ${\cal C}(\rho)$. Using the weak field approximations, we show that the gravitational effects decrease the maximum value of $K_{3}$ for some intervals of energy in such a way that there is no violation, while it leaves the maximum amount of the quantum coherence, ${\cal C}(\rho)$, unchanged.

\end{abstract}
\newpage
\newpage
\section{Introduction}

Gravitation and quantum mechanics usually do not play an important role simultaneously in low energy physical phenomena. However, at 1975 for the first time, Collela {\it et al.} proposed an experiment in which the results depended on both the gravitational and Planck constant \cite{collela}. This effect could be explained by considering a Newtonian potential inserted in the Schrodinger equation. Soon after, using semiclassical approximation, Stodolsky obtained the quantum mechanical phase of a free particle propagating in an external gravitational field in the weak field limit and investigated the gravitational effects by matter and light wave interferometry  \cite{stodolsky}. In addition to the subject of interferometry for neutrons and photons, the other place in which the gravitational effects and quantum mechanics coexist, is the context of neutrino oscillations.
Neutrinos are known to be the most fascinating particles of the standard model, since they can be utilized as special probes to study physics from quantum scales up to celestial scales. In fact, they have the ability to pass through ordinary matter with minimum interactions because they can only participate in weak interactions and their masses are extremely tiny to experience any reasonable gravitational influence. Nevertheless,  the effects of gravitation on neutrino oscillation have been investigated through calculating the change in quantum phase due to neutrino propagation in curved space-time \cite{guinti1997, guinti2004,neutrino oscillation in extended}.  

   On the other hand, quantum correlations are known as one of the cornerstones of the foundation of quantum information sciences and quantum technologies, since we can consider them as resource theories in quantum communications. In addition to their technological applications, one can use quantum correlations to study fundamental physical effects such as gravity at quantum mechanical scales. For instance, the gravitational effects on CHSH inequality which is known as one of the forms of the Bell inequalities have been studied \cite{quantum nonlocality in extended}. Also it has been investigated that the circular motion of particles in a Schwarzschild metric causes a decrease in the degree of violation of Bell's inequality \cite{epr correlation in gr}. In a similar study, the effects of a curved space-time described by the Kerr-Newman metric on the EPR correlations have been investigated \cite{epr correlation in kerr}.
   
    One of the other subjects in physics that neutrinos are used to study, is quantum information which is based mostly on quantum correlations such as entanglement and quantum coherence \cite{our article, blasone entanglement, blasone concurrence, quantifying coherence, quantum correlation alok, quantum information, ming, lgi violation in neutrino oscillation, lgi2017, lgi2019}. They can usually be expressed in terms of transition amplitudes and hence, oscillation probabilities \cite{quantum correlation alok}. Among the several researches on quantum correlations examining the quantumness in neutrino oscillations, are Refs. \cite{lgi violation in neutrino oscillation, lgi2019} that characterize the Leggett-Garg inequality for neutrino oscillation. Nonlocality in neutrino oscillation via  Bell type inequalities, i.e. the spatial counterparts of Leggett-Garg inequalities has also been studied \cite{ming}. Moreover, quantum effects such as entanglement and quantum coherence which seem to be distinct resource theories \cite{measuring coherence with entanglement, plenio2017}, have been investigated in the context of neutrino oscillation by Blassone {\it et al.} \cite{blasone concurrence} and  Xue-Ke Song {\it et al.} \cite{quantifying coherence}. Though it has been shown that these two quantum correlations, in the case of neutrino oscillations are not independent of each other \cite{our article}. In the Refs.  \cite{our article, quantifying coherence} quantum coherence has been treated quantitatively by defining the $l_{1}-$norm  as a coherence measure, which assigns a number to neutrino states. The $l_1$-norm is described as the summation over absolute values of all the off-diagonal elements of the corresponding density matrix. For neutrinos in two flavors, it can be written as $ \cal C(\rho)$=2 $\vert\sqrt{P_{osc}P_{sur}}\vert$ in terms of  the oscillation and survival probabilities \cite{our article}. 
   
  In fact, quantum coherence as a quantum correlation, can be investigated in the two qualitative and quantitative approaches; Leggett-Garg inequality (LGI) is a qualitative criterion and $l_1$-norm is a quantitative measure. In the case of the former, we should clarify that the violation of the LGI can be interpreted as appearing quantumness because quantum is the only theory that contradicts the postulates of the LGI.  Moreover, this quantumness is revealed if the states on which the measurements are carried out, are a coherent superposition of Hamiltonian eigenstates. Therefore, when we have treated the LGI by a quantum state which does not give violation, this means the quantum state does not have enough coherence. In this paper, we aim to study the gravitational effects on quantum coherence in neutrino oscillation by the two approaches mentioned above.  We consider the propagation of the neutrino in a Schwarzschild background for two flavors via the plane wave approach. Indeed, we use the method in which the quantum mechanical phase of neutrino is obtained by using the semiclassical approximation. With this assumption, the gravitational field is treated classically and it differs from the stochastic interactions between neutrinos and virtual black holes, which are due to fluctuations in space-time itself if gravity is a quantum force  \cite{stochastic perturbations}. 
  
  	In the next section, we will briefly review the neutrino oscillation in flat and curved space-time . In Sec. \ref{3}, we will study LGI in radial propagation of neutrinos in a Schwarzschild metric by calculating the $K_{3}$ parameter for neutrinos propagating outwards and inwards in a gravitational field. In Sec. \ref{4}, we will compute the $l_{1}$-norm for neutrino oscillation in two flavor framework via plane wave approach in a Schwarzschild metric in order to study its variations with respect to the proper distance $L_{p}$, when the neutrino is propagating towards or away from the gravitational source. Finally in Sec. \ref{5}, we will discuss our results and make the conclusion.

\section{Neutrino Oscillation in Flat and Curved Space-Time }\label{2}
In this section, we will briefly review the context of neutrino oscillation in flat and also curved space time. The state of a neutrino produced in a space-time point $A(t_{A},\vec{x}_{A})$ is obtained from the superposition of mass eigenstates, $ \vert\nu_{k}\rangle$,

	\be
	\vert\nu_{\alpha}\rangle=\sum_{k}U^{\ast}_{\alpha k}\vert\nu_{k}\rangle
	\label{flavor and mass eigenstates} 
	\ee   
	in which U is the unitary mixing matrix and in case of the two flavors it is shown as
\be
U=\left(\begin{array}{cc}
\cos(\theta)\  \sin(\theta)
\\ -\sin(\theta)\ \cos(\theta)
\end{array}\right)
\label{mixing matrix}
\ee

 This state as a weak interaction eigenstate is called the flavor eigenstate. There exists a discrepancy in energies, momenta and masses of different mass eigenstates $\vert\nu_{k}\rangle $ which leads them to propagate differently and hence, the neutrino oscillation takes place. In fact, the neutrino oscillation exhibits a relative shift in mass eigenstate phases when they arrive at the detector situated at the space-time point $B(t_{B},\vec{x}_{B})$. Actually, while the plane wave approach is adopted, the propagated mass eigenstates $\vert\nu_{k}\rangle $ are described as
\be
  \vert\nu_{k}(t, \vec{x})\rangle = e^{-i{\Phi_{k}}}\vert\nu_{k}\rangle 
\label{mass eigenstate propagation},
\ee 
where $\Phi_{k} (k=1,2) $ denoting the relative phase shift acquired by the mass eigenstates during their propagation, is given by
\be
\Phi_{k}=E_{k}(t_{B}-t_{A})-\vec{p_{k}}.(\vec{x}_{B}-\vec{x}_{A})
\label{flat phase1}.
\ee
In general, in the case of two flavors, the probability of detecting a neutrino as $\vert\nu_{\mu}\rangle $ that is produced as $ \vert\nu_{e}\rangle $ is given by
	\be
	P(\nu_{e}\rightarrow\nu_{\mu})=\vert\langle\nu_{\mu}\vert\nu_{e}(t_{B}, \vec{x}_{B})\rangle\vert^{2}=\sin ^{2}(2\theta)\sin ^{2}(\dfrac{\Phi_{jk}}{2}),
	\label{transition probability}
\ee		
in which, $\Phi_{jk}=\Phi_{j}-\Phi_{k}$. Of course, in order to have the neutrino oscillation, the difference of the phase shift must not lead to the decoherence of the flavor eigenstates due to the propagation. In other words, the propagation length has to be smaller than the coherence length. 
	
	In flat space-time described by the Minkowski metric, applying the ultra relativistic limit of $ (t_{B}-t_{A})\simeq \vert x_{B}-x_{A}\vert $, together with the relativistic expansion $m_{k}\ll E_{k}$, one can write  
		\be 
	\Phi_{jk}\simeq \dfrac{\Delta m^{2}_{jk}}{2E_{0}}\vert x_{B}-x_{A}\vert
	\label{flat phase2}.
	\ee
Here, $\theta$ is the mixing angle, $\Delta  m_{jk}^{2} $ is the mass squared difference and $E_{0}$ is the energy for the massless neutrino measured by the observer at infinity.

	Now, to generalize our discussion to a curved space-time, we can replace the quantum phase given by the Eq. (\ref{flat phase1}) with its covariant form
\be
\Phi_{k}=\int_{A}^{B}p_{\mu}^{(k)}dx^{\mu} 
\label{covaiant phase},
\ee
where, $p_{\mu}^{(k)}$ is the canonical conjugate momentum to the coordinate $x^{\mu}$ and is given by 
\be
p_{\mu}^{(k)}=m_{k} g_{\mu\nu}\dfrac{dx^{\nu}}{ds}
\label{momentum}.
\ee
In above equation, $g_{\mu\nu}$ represents the metric tensor and $ds$ is the line element. For instance, let us consider the Schwarzschild metric
  \be
  ds^{2}=B(r)dt^{2}-B(r)^{-1}dr^{2}-r^{2}d\theta^{2}-r^{2}\sin^{2}\theta d\phi^{2},
	\label{Schwarzschild line element}
\ee
where 
\be
B(r)=(1-\dfrac{2GM}{r}),
\label{B(r)}
\ee
in which, $G$ is the Newtonian constant and $M$ stands for mass of the gravitational source \cite{weinberg}. In this metric, the gravitational field is isotropic, therefore, we may consider the propagation of neutrinos in the equatorial plane $\theta=\pi/2$. In this case, the phase obtained by the mass eigenstates during the neutrino propagation from the source at the space-time point $A(t_{A}, r_{A}, \phi_{A})$ to the detector at the space-time point $B(t_{B},r_{B},\phi_{B})$, may be written as 
\be
\Phi_{k}=\int_{A}^{B}[E_{k}dt-p_{k}(r)dr-J_{k}d\phi]
\label{schwarzschild phase}.
\ee
Here, $E_{k}\equiv p_{t}^{(k)}$, $p_{r}\equiv -p_{r}^{(k)}$ and $J_{k}(r)\equiv -p_{\phi}^{(k)}$, are the components of the canonical momentum $p_{\mu}^{(k)}$ \cite{hobson}. 
For convenience, in presence of the gravitational effects, it is better to consider the neutrino propagation over the proper distance $ L_{p}$, which is  generally defined by the relation
	\be
\begin{array}{cc}
L_{p}\equiv\int^{r_{B}}_{r_{A}}\sqrt{g_{rr}}\ dr
\\
\\ = r_{B}\sqrt{1-\dfrac{2GM}{r_{B}}}-r_{A}\sqrt{1-\dfrac{2GM}{r_{A}}}
\\
\\+2GM\left[ \ln\left(\sqrt{r_{B}-2GM}+\sqrt{r_{B}}\right) -\ln \left( \sqrt{r_{A}-2GM}+\sqrt{r_{A}}\right) \right]

	\label{proper distance}. 
\end{array}
	\ee

Assuming the weak field approximation, one can easily calculate $L_{p} $ to be 
\be
L_{p}\simeq r_{B}-r_{A}+GM \ln\dfrac{r_{B}}{r_{A}},
\label{proper length in weak field approximation}
  \ee
  where $ r_{A}$ and $ r_{B}$ are, respectively, the positions of the source and the detector which are measured with respect to the gravitational source reference frame.	 In addition, we should note that using the ultra relativistic limit, one can write the relation between proper time interval and proper distance as
\be
L_{p}=c\tau
\label{proper distance and time}
\ee	
in which $ \tau$ is the proper time.

	At this point, we restrict our discussion to the motion of neutrinos in a Schwarzschild gravitational field in radial propagation. For radial propagation of neutrinos in a Schwarzschild gravitational field, i.e., $ d\phi=0 $ in Eq.  (\ref{schwarzschild phase}),  there is no angular momentum. As for the flat-space time, the relativistic expansion $m_{k}\ll E_{k}$ can be applied. Thus, assuming $ 0 <B(r)\leq 1$, after some calculations we have \cite{guinti1997}
\be
\phi_{k}\simeq \int_{r_{A}}^{r_{B}}\dfrac{m_{k}^{2}}{2E_{0}}dr
\label{radial1},
\ee	
which is integrated along the light-ray trajectory to be
\be 
\Phi_{k}\simeq \dfrac{m_{k}^{2}}{2E_{0}}\vert r_{B}-r_{A}\vert
\label{radial2}.
\ee
It should be noticed that $ \vert r_{B}-r_{A}\vert$ is the coordinate difference and as mentioned earlier, in the curved space-time it can be written in terms of the proper distance. Here, the energy $E_{0}$ is the energy of the massless neutrino that is measured by the observer located at $r=\infty$. However, it may be convenient to rewrite the oscillation phase defined in Eq.(16) in terms of the local energy, denoted by ${E_{0}^{loc} (r_{B})}$, measured by the observer at the detector at $r_{B}$. This local energy is related to the $E_{0}$ through the relation $E_{0}^{loc}(r)=\vert g_{tt}\vert^{-1/2}E_{0} $ \cite{wheeler}. Thus regarding the Eq. (\ref{Schwarzschild line element}), we can write
	\be
	E_{0}^{loc}(r_{B})=(1+\dfrac{GM}{r_{B}})E_{0}.
	\label{local and initial energy}
	\ee
Finally, using Eqs.(\ref{proper length in weak field approximation}) and (\ref{local and initial energy}) we can express the Eq.(\ref{radial2}) as follows \cite{guinti1997} 

\be
\Phi_{kj}\simeq(\dfrac{\Delta m^{2}_{kj}L_{p}}{2E_{0}^{loc} (r_{B})})\left[  1-GM(\dfrac{1}{L_{p}}\ln \dfrac{r_{B}}{r_{A}}-\dfrac{1}{r_{B}}) \right].   
\label{oscillation phase}
\ee

   In this equation, the corrections due to the gravitational effects are evident from the second term in the square parentheses. In the following, we wish to study the effects of gravitation on the LGI violation as well as the amount of quantum coherence ($l_{1}$-norm). In order to have an appropriate evaluation, we take $\Delta m^{2}_{12}=7.92\times 10^{-5} eV^{2}$, $ \theta_{12}=0.59 $ and  $ GM=3\times 10^{7}\mathrm{Km}$, which can be the Schwarzschild radius of a supermassive black hole.

\section {LGI in Radial Propagation of Neutrino in a Schwarzschild Metric}  \label{3}

In this section, we intend to study the gravitational effects on the LGI using the time correlations in neutrino oscillations. 
 In fact, the violation of the LGI is a confirmation of the existence of quantum coherence. The LGI is based on the two main postulates; macroscopic realism and noninvasive measurability \cite{leggett-garg, realism, lgi review}. The first postulate means that the measurements performed on a system, will only reveal the values that already exist. 
  The second postulate implies that the noninvasive measurements can be carried out on the system without changing the state of the system. The simplest form of LGI is constructed in terms of the parameter $K_{3}$ which is defined as
 	\be
K_{3}= C(t_{1},t_{2})+C(t_{2}, t_{3})-C(t_{1},t_{3}), \label{leggett-garg1}
\ee 
in which $C(t_{i}, t_{j})= \left \langle \hat{Q}(t_{i}) \hat{Q}(t_{i}) 
 \right\rangle$ and $\hat{Q}(t)$ is a dichotomic variable that can take only one of two discrete values which are usually labeled by convention $+1$ and $-1$ \cite{lgi review}.
	According to LGI, this parameter should be smaller than one, otherwise its assumptions are violated. We should notice that in order to check the LGI inequality, we need to use a physical state which is a coherent superposition of some eigenstates to calculate the corresponding correlations. Of course, this feature can occur in quantum mechanics, while this theory is in conflict with the postulates on which the LGI is constructed. So, if this inequality is violated, it means that we have a phenomenon that is inconsistent with the corresponding postulates. On the other hand, if we choose a quantum state for which this inequality is not violated, it could mean that the assumed state is not coherent enough, therefore, for example, $K_3$ does not exceed unity.
	
 If one wants to investigate the validity of the LGI postulates, it is necessary to provide appropriate conditions that ensure compliance with the LGI postulates. However, there are some experimental difficulties satisfying the noninvasive measurement. Furthermore, its postulate is in conflict with quantum mechanics and it is natural to find a quantum mechanical set up that violates the LGI. Meanwhile, the LGI may be derived instead
under the assumption of stationarity \cite{stationarity}. Accordingly, the
correlation functions $C_{ij}$ depend only on the time difference
$\tau \equiv t_j-t_i$  between various measurements.
 There are several studies based on the \textit{stationarity} condition, in which the LGI has been investigated using the neutrino oscillation \cite{lgi violation in neutrino oscillation, lgi2017}. Nevertheless, since we take account of the gravitational effects on LGI in the present study, the condition of \textit{stationarity} for neutrino oscillation  is not necessarily satisfied. Therefore, we expect the LGI violation by neutrino oscillation with the gravitational modifications unless the coherent superposition is deflected. Hence, we can take the violation of the LGI as a qualitative criterion for quantum coherence.
 The LGI for neutrinos without assuming the condition of \textit{stationarity} is investigated in Ref. \cite{lgi2019}. We must also write the parameter $K_{3}$ to be Lorentz invariant. Thus, one needs to rewrite the  time correlation functions in Eq. (\ref{leggett-garg1}) in terms of the proper time $\tau$ and, for simplicity, we adopt the equal proper time intervals ($ \tau_{3}-\tau_{2}=\tau_{2}-\tau_{1}=\tau$). Furthermore, according to the ultra relativistic limit introduced in Eq. (\ref{proper distance and time}), we can write the parameter $K_{3}$ in Eq.(\ref{leggett-garg1}) in terms of the proper distance $L_{p}$ as follows:

\be
K_{3}= C(0, L_{p})+C(L_{p}, 2L_{p})-C(0, 2L_{p}) \label{leggett-garg}.
\ee 
Here, in the case of neutrino oscillation, the observable for which the correlation functions are evaluated is $\hat{Q}=2\vert\nu_{\alpha}\rangle\langle\nu_{\alpha}\vert-1 $. If the neutrino is still in its initial flavor state the outcome is $+1$ and otherwise it will be $-1$.

Now let us consider a neutrino source located at the radial distance $ r_{A}$ from the center of the gravitational field of a nonrotating object with a spherical symmetry that is described by the Schwarzschild metric. We intend to construct the parameter $ K_{3}$ between the source and the two hypothetical detectors that are placed in radial distances $r_{B1}$ and $r_{B2}$ from the center of the gravitational source. We consider the case in which the initial neutrino flavor state is $\vert \nu_{\mu}\rangle$. In general, the correlation of operator $\hat{Q}$ between two different proper distances is defined by \begin{equation}
	C(L_{p}^{(1)},L_{p}^{(2)})= \langle\mu\vert \frac{1}{2}\{\hat{Q}(L_{p}^{(1)}),\hat{Q}(L_{p}^{(2)})\}\vert \mu\rangle,
\end{equation} where $\hat{Q}(L_p^{(i)})=U^\dagger(L_p^{(i)})\hat{Q}U(L_p^{(i)})$. Superscript $(i)$ refers to either the source or one of the detectors. Here, $U(L_{p})$ is the time evolution matrix in the flavor space, whose elements are defined as
\be
U_{\delta \lambda}(L_{p})= \sum_{k}U^{\ast}_{\delta k} e^{-i\phi_{k}(L_{p})}U_{\lambda k}.
\label{transition amplitude}
\ee
In fact, $ U_{\delta \lambda}(L_{p})$ is the transition amplitude. We should note that the time evolution operator in Eq.(\ref{transition amplitude}) is unitary when using Eq.(\ref{radial2}); i.e., one gives the $\phi_{k}$ in terms of the $r_{B}$. However, in Eq. (\ref{transition amplitude}), we have written  $r_{B}$ in terms of the $L_{p}$ using Eq. (\ref{proper length in weak field approximation}) and thus $U(L_p)$ is not unitary in terms of $L_p$.

Thus we have for the three correlations in Eq. (\ref{leggett-garg})
\begin{eqnarray}
C(0,\kappa L_{p})=2P_{\nu_{\mu}\rightarrow \nu_{\mu}}(\kappa L_{p})-1,\hspace{1 cm}           \text{for} \           \alpha =\mu\nonumber
\\ \hspace{1.9 cm}=1-2P_{\nu_{\mu}\rightarrow \nu_{\alpha}}(\kappa L_{P}),\hspace{1 cm}               \text{for} \          \alpha \neq \mu
\end{eqnarray}
in which $\kappa=1,2$ and 
\begin{eqnarray}
 C(L_{p}, 2L_{p})=&&\hspace{-0.6 cm}(1/2) \left[ \langle\nu_{\mu}\right\vert 4U^{\dagger}(L_{p})\vert \nu_{\alpha}\rangle \langle\nu_{\alpha}\vert U(L_{p}) U^{\dagger}(2L_{p})\vert\nu_{\alpha}\rangle\langle\nu_{\alpha}\vert U(2L_{p})\vert\nu_{\mu}\rangle\nonumber\\
  &&\hspace{0.4cm}+\langle\nu_{\mu}\vert 4U^{\dagger}(2L_{p})\vert\nu_{\alpha}\rangle\langle\nu_{\alpha}\vert U(2L_{p})U^{\dagger}(L_{p})\vert\nu_{\alpha}\rangle\langle\nu_{\alpha}\vert U(L_{p})\vert\nu_{\mu}\rangle\nonumber\\
  &&\hspace{0.4cm}-\langle\nu_{\mu}\vert 4U^{\dagger}(L_{p})\vert\nu_{\alpha}\rangle\langle\nu_{\alpha}\vert U(L_{p})\vert \nu_{\mu}\rangle\nonumber\\
   &&\hspace{0.4cm}-\langle\nu_{\mu}\vert 4U^{\dagger}(2L{p})\vert\nu_{\alpha}\rangle\langle\nu_{\alpha}\vert  U(2L_{p})\vert\nu_{\mu}\rangle +2 ]. 
\label{1212}
\end{eqnarray}
One can simplify the recent equation as follows:
\begin{eqnarray}
	C(L_{p}, 2L_{p})=&&\hspace{-0.6 cm}4\bar{U}_{\mu \alpha}(L_{p})U_{\alpha \mu}(L_{p})\bar{U}_{\mu \alpha}(2L_{p})U_{\alpha \mu}(2L_{p})\nonumber\\
	&&\hspace{-0.6 cm}+2\bar{U}_{\mu \alpha}(L_{p})U_{\alpha e}(L_{p})\bar{U}_{e \alpha}(2L_{p})U_{\alpha \mu}(2L_{p})\nonumber\\
	&&\hspace{-0.6 cm}+2\bar{U}_{\mu \alpha}(2L_{p})U_{\alpha e}(2L_{p})\bar{U}_{e \alpha}(L_{p})U_{\alpha \mu}(L_{p})\nonumber\\
	&&\hspace{-0.6 cm}-2P_{\nu_{\mu}\rightarrow \nu_{\alpha}}(L_{p})-2P_{\nu_{\mu}\rightarrow \nu_{\alpha}}(2L_{p})+1.
\end{eqnarray}
 Consequently, we obtain the following expression for $K_{3}$ in terms of the probabilities and the transition matrix elements:
\begin{eqnarray}
K_{3}=&&\hspace{-0.6cm}1-4P_{\nu_{\mu}\rightarrow \nu_{e}}(L_{p})+4\bar{U}_{\mu e}(L_{p})U_{e \mu}(L_{p})\bar{U}_{\mu e}(2L_{p})U_{e \mu}(2L_{p})\nonumber
\\ &&\hspace{-0.6cm}+4Re[\bar{U}_{\mu e}(2L_{p})U_{e e}(L_{p}){U}_{e e}(2L_{p})\bar{U}_{e \mu}(L_{p})]\hspace{1cm} \text{for}\  \alpha =e,
\label{parameter k3}
\end{eqnarray}
and
\begin{eqnarray}
K_{3}=&&\hspace{-0.6cm}1-4P_{\nu_{\mu}\rightarrow \nu_{\mu}}(2L_{p})+4P_{\nu_{\mu}\rightarrow \nu_{\mu}}(L_{p})
 P_{\nu_{\mu}\rightarrow \nu_{\mu}}(2L_{p})\nonumber \\
 &&\hspace{-0.6cm}+4Re[\bar{U}_{\mu e}(2L_{p})U_{\mu \mu}(L_{p}){U}_{\mu \mu}(2L_{p})\bar{U}_{e \mu}(L_{p})]
  \hspace{1cm}    \text{for}\     \alpha =\mu.
\label{alpha=mu parameter k3}
\end{eqnarray}

	We preform our calculations in the weak field approximation, since in the distance scales we intend to work in, the weak field approximation holds. Albeit, we should write our relations in terms of the local energy $E^{\text{loc}}_{0}(r_{B})$ measured by the observers situated at the two detectors, since we intend to study the behavior of $K_{3}$ with respect to variations of an unique energy, we rewrite both energies measured by the two detectors in terms of the energy $E_{0}$ measured by the observer at the infinity. To better enlighten the gravitational effects on the parameter $K_{3}$, we will evaluate and compare the parameters $K_{3}$ for the two different cases of the neutrinos propagating radially outwards and towards the gravitational source.

\begin{itemize}
\item In the case of neutrinos propagating radially outwards the gravitational source, we can write the distances $r_{B1} $ and $r_{B2}$ from the Eq. (\ref{proper length in weak field approximation}) as follows:
	\be    
	r_{B1}=L_{p}+r_{A}-GM \ln (\dfrac{L_{p}}{r_{A}}+1),
		\label{rB1}
	\ee
\be
r_{B2}=2L_{p}+r_{A}-GM \ln (\dfrac{2L_{p}}{r_{A}}+1).
\label{r_{B2}}
\ee
Therefore, up to the first order of the weak field approximation, ${\cal O}(GM/r) $, we can express the oscillation phases as
\be
\Phi_{kj}(L_{p})\simeq \dfrac{\Delta m^{2}_{kj}L_{p}}{2E_{0} }\left[ 1-GM \left( \dfrac{1}{L_{p}} \ln ( \dfrac{L_{p}}{r_{A}}+1)\right) \right],
\label{phi outwards11}
\ee
and
\be
\Phi_{kj}(2L_{p})\simeq \dfrac{\Delta m^{2}_{kj}2L_{p}}{2E_{0} } \left[ 1-GM \left( \dfrac{1}{2L_{p}}\ln( \dfrac{2L_{p}}{r_{A}}+1) \right) \right].
\label{phi outwards22}
\ee

 By using these two recent relations and Eqs.(\ref{transition probability}) and (\ref{transition amplitude}) in Eqs. (\ref{parameter k3}) and (\ref{alpha=mu parameter k3}), the parameter $K_{3}$ can be constructed. 
In order to specify the gravitational effects on the parameter $K_{3}$ and to checkout the LGI violation, we take the relative parametric space such that the gravitational effects might be remarkable in neutrino oscillation. Hence, we choose the locations of the two detectors such that  the proper distance becomes $L_{p}\simeq 3\times10^{8}\mathrm{Km} $. The radial distance of the source is selected to be $r_{A}\simeq 10^{8}\mathrm{Km}$ and the energy order is taken about several hundred $\mathrm{TeV}$s. Maybe it should be enlightened that our choices of the parameters are not far from the outputs of the astrophysical models. Among the existing models describing the production of high energy neutrinos near a massive object \cite{galactic neutrino1,galactic neutrino2,galactic neutrino3,blackhole merger,high energy neutrino,high energy neutrino2}, there is a conjecture that would create a sort of physics case for our parameter choices \cite{high energy neutrino2}. According to this conjecture, the production of neutrinos with the energy scale of $\mathrm{TeV}$ can take place within a half of the	Schwarzschild radius of a supermassive black hole.
 Accordingly,  the variation of the parameter $K_{3}$ is plotted with respect to $ E_{0}$ in the interval $1.5\times 10^{2}$ to $5\times 10^{2}$ TeV in Fig. 1(a). For comparison, the parameter $K_{3}$ for the case of the flat space time, i.e., $M=0$ is also drawn. 

\

\begin{figure}[h!]
	\begin{centering}
		\subfigure[]{\includegraphics[scale=0.30]{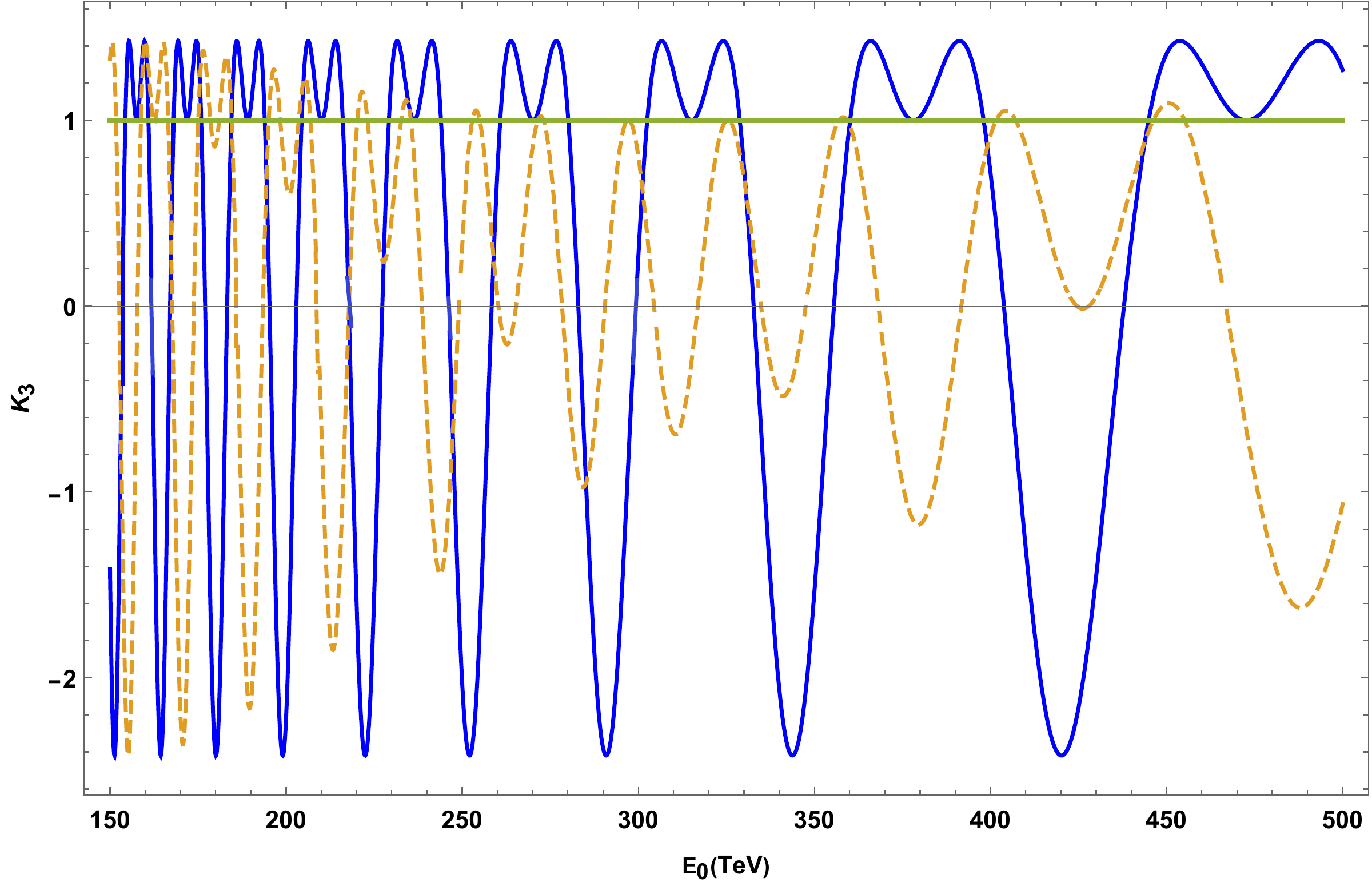}}\subfigure[]{\includegraphics[scale=0.25]{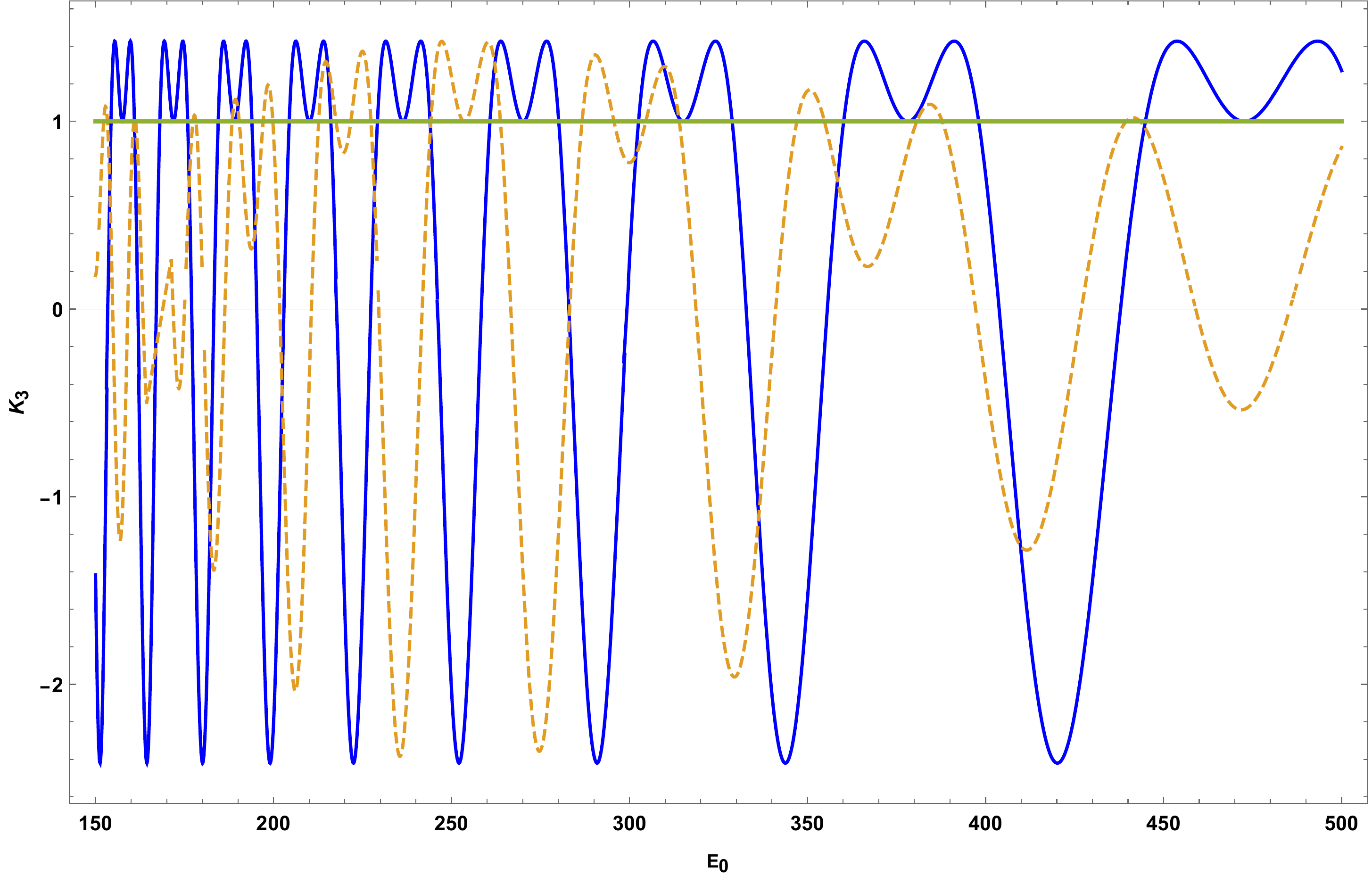}}
		\caption{Variations in $K_{3}$ as a function of the energy $E_{0}$ in TeV, for neutrinos radially propagating outwards (a) and inwards (b) the gravitational source. Blue (solid) and orange (dashed) curves represents the $K_{3}$ in flat and curved space time, respectively.} 
		\label{1}
	\end{centering}
\end{figure}

\item In the case of neutrino propagating radially towards the gravitational source, we place the neutrino production source at $\acute{r}_{A}$. The parameter $K _{3}$ is calculated for the neutrinos propagating from this point to the detectors situated at the places $\acute{r}_{B1}$ and $\acute{r}_{B2}$ given by the relations
\be 
\acute{r}_{B1}=\acute{r}_{A}-L_{p}-GM \ln (1-\dfrac{L_{p}}{\acute{r}_{A}}),
\label{rprime b1}
\ee
and
\be
\acute{r}_{B2}=\acute{r}_{A}-2L_{p}-GM \ln (1-\dfrac{2L_{p}}{\acute{r}_{A}}).
\label{rprime b2}
\ee
Consequently, we can again evaluate the oscillation phases as
\be
\acute{\Phi}_{kj}(L_{p})\simeq \dfrac{\Delta m^{2}_{kj}\acute{L}_{p}}{2E_{0} }\left[ 1+GM \left( \dfrac{1}{\acute{L}_{p}} \ln (1- \dfrac{\acute{L}_{p}}{\acute{r}_{A}})\right) \right],
\label{phi outwards1}
\ee
and
\be
\acute{\Phi}_{kj}(2\acute{L}_{p})\simeq \dfrac{\Delta m^{2}_{kj}2\acute{L}_{p}}{2E_{0} } \left[ 1+GM \left( \dfrac{1}{2\acute{L}_{p}}\ln( 1-\dfrac{2\acute{L}_{p}}{\acute{r}_{A}}) \right) \right].
\label{phi outwards2}
\ee
Similar to the case of neutrinos propagating outwards, we can plot the parameter $K_{3}$ as shown in the Fig. 1(b), by considering $\acute{r}_{A}=6.5\times 10^8 Km$ and $\acute{L}_{p}=3\times 10^8 Km$.

\end{itemize}

The following two points are obvious from the two diagrams in Fig.1
	\begin{itemize}

	\item
	   It is seen that due to the presence of gravitational effects, there exists a damping in the maximum value of the parameter $K_3$ in some intervals of energy such that the LGI violation diminishes for these intervals. 
	  Indeed, as was said, the violation of the LGI can be interpreted as appearing as quantumness because quantum is the only theory that contradicts the principles of the LGI. This quantumness is revealed if the states on which the measurements are carried out, are a coherent superposition of Hamiltonian eigenstates. When we have treated the LGI by a quantum state which does not give violation, it means that this quantum state does not possess enough coherence. Therefore, the LGI investigation shows the decrease in coherence due to the gravitational effects.
	 
		\item 

	As another interesting result, we see that the gravitational effects lead to the occurance of a phase shift in the value of the parameter $K_3$ in comparison to the corresponding one in the flat space-time. This phase shift depends on both energy and the relative proper distances.
	So, there exists some range of the parameter by which the value of  $K_{3}$ in curved space-time is larger than the one in the flat space-time. Of course, we know that even in presence of the quantum coherence, it is not the case that we will be able to see the LGI violations for any choice of parametric space. Rather, if there is enough coherence, one can find a region of parametric space in which this inequality violates. Therefore, according to the results obtained from this study, the gravitational effects cause changes in the region of the parametric space suitable for violating LGI.
	The physical justification is as follows: we describe neutrinos by plane waves in this study and the corresponding phases are modified by the gravitational effects. Hence, when we encounter the LGI violation, this means that the phase of the distinguished terms constructing $K_{3}$ is such that they are summed constructively and otherwise destructively. 

	\end{itemize}

\section{Calculation of $l_{1}$-norm for Neutrino Oscillation in Schwarzchild Metric}\label{4}
 
 After investigating the gravitational effects on the quantumness of neutrino oscillation via calculating the parameter $K_{3}$ for the two different cases described in the last section, we wish to study quantitatively the quantum coherence in neutrino propagation in presence of the gravitational effects. As was said in the introduction, the amount of quantum coherence can be obtained for the neutrino oscillation by calculating the $l_{1}$-norm \cite {our article, quantifying coherence}.  In general, this quantity is defined as \cite{plenio2014} :
\be
{\cal C}(\rho)=\sum_{i\neq j}\vert\rho_{ij}\vert\geq 0 \label{coherence measure}
\ee
where the summation is over the absolute values of all the off-diagonal elements $\rho_{ij}$  of a corresponding density matrix $ \rho $. The maximum possible value for ${\cal C}(\rho) $ is bounded by $ {\cal C}_{max}=d-1 $, where d is the dimension of the density matrix $ \rho $. Here we consider the two flavor neutrino, thus the maximum value of $l_{1}$-norm will be 1.
 The $l_{1}$-norm can be expressed in terms of the transition and survival probabilities between different flavor modes written in terms of the proper distance $L_{p}$. Therefore, if the initial flavor state is $\vert\nu_{\mu}\rangle$ we have
\be
{\cal C}_{\mu}= 2(\sqrt{P_{\mu e}(L_{p})P_{\mu\mu}(L_{p})}).
\label{cal}
 \ee

 We consider the gravitational effects in the weak field approximation for the Schwarzschild metric. The calculations are carried out for the two cases of the neutrinos propagating radially outwards and inwards. 
 Using the plane wave approach, one can obtain straightforwardly the transition probabilities as 
\begin{eqnarray}
P_{\nu_{\mu}\rightarrow \nu_{e}}(L_{p})=\sin^{2}(2\theta) \sin^{2}\Big[&&\hspace{-0.6cm} \frac{\Delta m^2_{12}L_p}{4E_{0}^{\text{loc}}}\nonumber\\
&&\hspace{-2.5cm}\times\Big\{ 1-GM \left( \dfrac{1}{L_{p}}\ln( \dfrac{L_{p}}{r_{A}}+1)-\dfrac{1}{L_{p}+r_{A}}  \right) \Big\}
\Big] 
\label{probability at lp},
\end{eqnarray}
and 
\begin{eqnarray}
\acute{P}_{\nu_{\mu}\rightarrow \nu_{e}}(\acute{L}_{p})=\sin^{2}(2\theta) \sin^{2}\Big[&&\hspace{-0.6cm} \dfrac{\Delta m^{2}_{12}\acute{L}_{p}}{4E_{0}^{\text{loc}}}\nonumber\\
&&\hspace{-2.5cm}\times\Big\{ 1+GM \left( \dfrac{1}{\acute{L}_{p}}\ln( 1-\dfrac{\acute{L}_{p}}{\acute{r}_{A}})+\dfrac{1}{\acute{L}_{p}+\acute{r}_{A}}  \right) \Big \}\Big] 
\label{probability at lpprime}.
\end{eqnarray}
The survival probabilities are also 
 \be
P_{\nu_{\mu}\rightarrow \nu_{\mu}}(L_{p})=1- P_{\nu_{\mu}\rightarrow \nu_{e}}(L_{p})
\label{survival probability at lp},
\ee
and 
\be
\acute{P}_{\nu_{\mu}\rightarrow \nu_{\mu}}(\acute{L}_{p})=1- \acute{P}_{\nu_{\mu}\rightarrow \nu_{e}}(\acute{L}_{p})
\label{survival probability at lpprime}.
\ee

  Here, in our calculations, $E^{loc}_{0}(r_{B})$ i.e., the energy of the massless neutrino measured by the observer at the detector situated at the radial distance $r_{B}$ from the center of the gravitational source is used. In order to plot the variations of ${\cal C}_{\mu}$ in Eq. (\ref{cal}) in terms of the proper distance $L_{p}$ we take $E^{loc}_{0}(r_{B})=3\times 10^{2}$ TeV. For neutrinos propagating outwards, we suppose $r_{A}=10^{8}$ Km and $2\times 10^{8}$ Km $\leq L_{p} \leq 4\times 10^{8}$Km and for neutrinos propagating inwards,  $\acute{r}_{A}=4\times10^{8}$ Km and $1.5\times 10^{8}$ Km $\leq \acute{L}_{p} \leq 3\times 10^{8}$Km. The corresponding diagrams are found in Fig. 2(a) and Fig. 2(b).
\begin{figure}[h!]
	\begin{center}
		\subfigure[]{\includegraphics[scale=0.46]{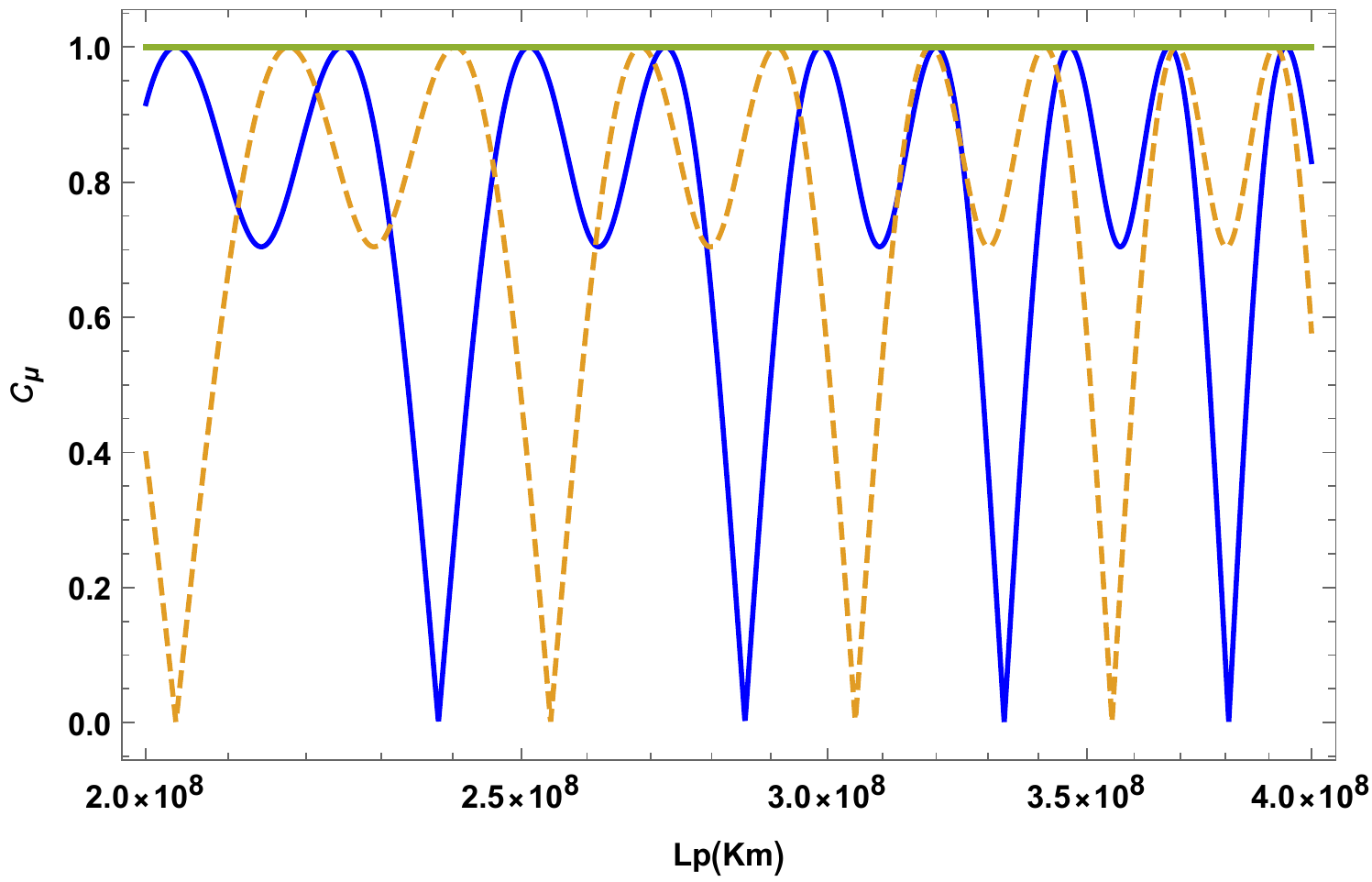}}\subfigure[]{\includegraphics[scale=0.44]{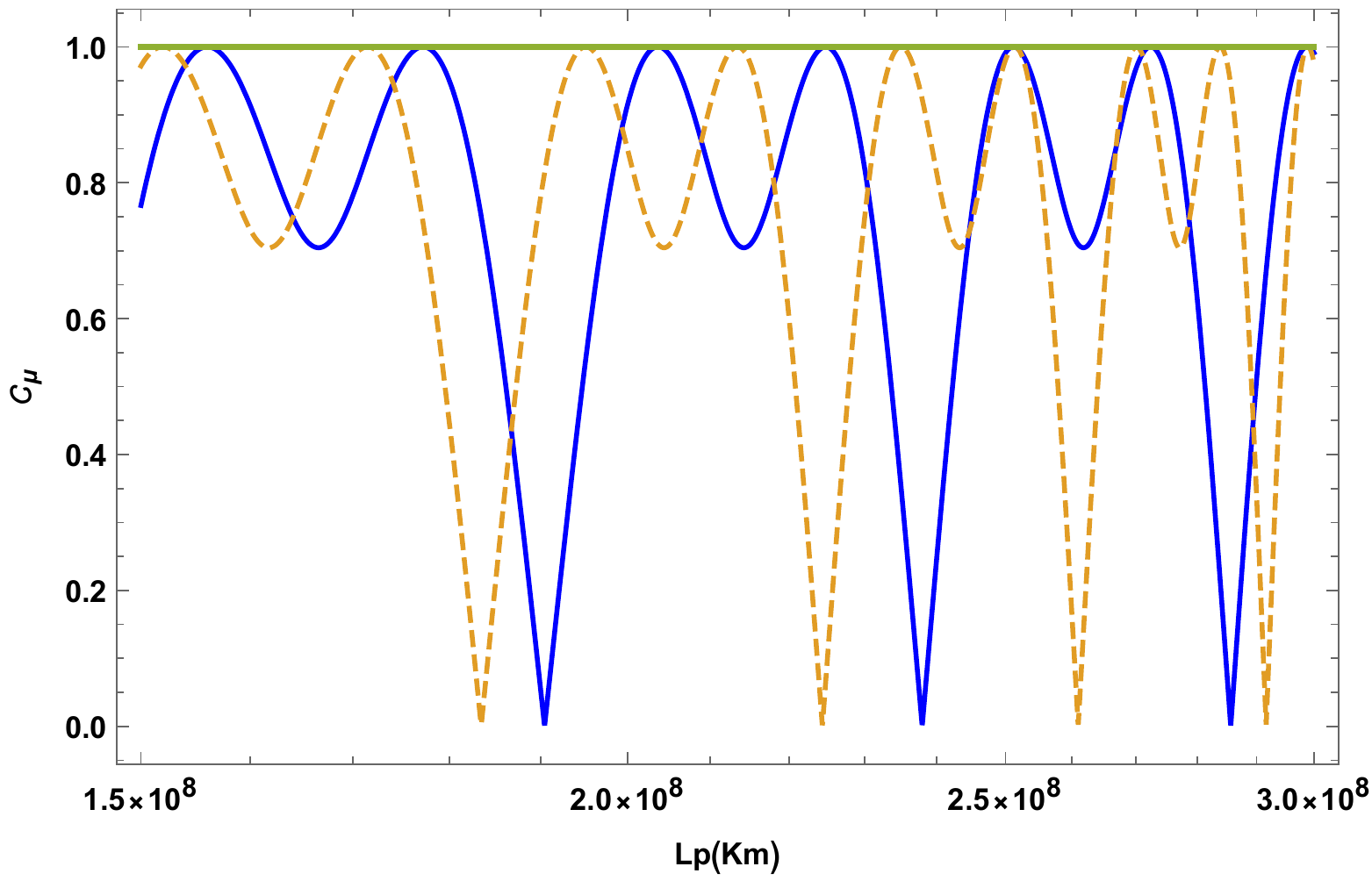}}
		\caption{Variations of ${\cal C}_{\mu}$ as a function of $L_{p}$, for neutrinos radially propagating (a) outwards and (b) inwards the gravitational source. Blue (solid) and orange (dashed) curves represent the ${\cal C}_{\mu}$ in the flat and curved space time, respectively.} 
		\label{2}
	\end{center}
\end{figure}

It is notable from the two diagrams in Fig. 2 that although as we expect, the wave length shows local dependence in both outwards and inwards propagation, we have no decrease in the maximum value of the quantum coherence.

\section{Conclusion}\label{5}

One of the important questions in physics is about the quantum effects of the gravitation, either we have a quantum gravity or the gravitation can play any role at the quantum scales.
The most important quantum aspect that plays role in neutrino oscillation is the quantum coherence. In this paper, we have studied the gravitational effects on quantum coherence in neutrino oscillation via both qualitative (through investigating the violation of LGI) and quantitative (by calculating $l_{1}$-norm) manners. 

 In case of the former, we have rewritten the parameter $K_{3}$ that is known as the simplest form of the LGI, in terms of the proper distance $L_{p}$. Next we constructed the parameter $K_{3}$ for two flavor neutrinos propagating radially outwards and inwards between a neutrino source and the two detectors placed in a Schwarzschild metric. Then we have given a numerical example for the variations of the parameter $K_{3}$ with respect to the energy $E_{0}$ in the weak field approximation. The results are demonstrated by Fig. 1. We concluded that the gravitational effects decrease the maximum value of the parameter $K_{3}$ for some energy intervals. More explicitly, there exists some energy intervals for which we have no LGI violation for neutrinos propagating in curved space-time but we do have the LGI violation in flat space-time. 
In case of the latter, we have calculated the $l_{1}$-norm, ${\cal C}(\rho)$,  for the above physical situation. We reckon that although a local dependence in wave lengths of the oscillation is observable in ${\cal C}(\rho)$ when the neutrinos propagate in curved space-time, the maximum amount of the ${\cal C}(\rho)$ remains unchanged (please see Fig. 2.).

 	  In this study, the considered setup seems to be a gedanken experiment and the distances chosen for the LGI experiment are somewhat exaggerated.
 	  However, the issues presented in this study may be considerable for future experiments which may be designed to examine the decoherence effects for long baseline experiments by the messaging particles other than neutrinos between satellites at the vicinity of the Earth. Recently, some efforts have been made to create a quantum network in space \cite{spacenetwork1,spacenetwork2,spacenetwork3}. Therefore, it can be helpful to study the various quantum resources such as quantum coherence in the curved space-time.
 	  
 	  Furthermore, we have treated neutrinos by plane wave and the decoherence effects due to the separation of the mass eigenstate wave packets (for instance see Ref. \cite{wp}) have been ignored. The wave packet decoherence effects in the presence of the strong gravitational fields have been studied in Ref. \cite{wpsf}. However, the plane wave assumption does not make our discussion inaccurate because modifications due to neutrino localization (wave packet approach) only play an important role when the propagation length is of the order of the coherence one. Therefore, we do not see any damping in the maximum value of $l_1$-norm. But under the same condition, we have a damping in the LGI which is due to the constructive and destructive effects in some terms constructing $K_{3}$. Thus, this statement shows that the LGI violation and $l_1$-norm are independent measure criteria of coherence. This reminds us of some mixed states such as the Werner state \cite{werner}, in which there are regions where we
 have no CHSH violation and, therefore, no nonlocality, but still get a nonzero value for the measure quantities of the entanglement.

In general, there exists another scenario for decoherence of neutrino states in vacuum; the effects caused by the real and microscopic virtual black holes, speculated from the quantum gravity considerations, can lead to the decoherence of neutrino states (for instance see \cite{stochastic perturbations}). These theories could have an impressive role on the neutrino oscillation but there is no empirical indication for them and they are bounded via various studies performed in this field \cite{elis,liu,blennow1,blennow2}. Meanwhile, considering such a hypothesis is a new issue that goes beyond the scope of the current article.

\end{document}